\documentclass[aps,prc,12pt,onecolumn,amsmath,amssymb,superscriptaddress]{revtex4-2}
\usepackage{graphicx}
\usepackage{color}
\usepackage{lineno}

\newcommand{\be}{\begin{equation}}
\newcommand{\ee}{\end{equation}}
\newcommand{\beq}{\begin{eqnarray}}
\newcommand{\eeq}{\end{eqnarray}}
\newcommand{\ba}{\begin{array}}
\newcommand{\ea}{\end{array}}

\begin{document}

\title{CP Violation Problem}

\vspace{3mm}
\date{\today}

\author{\mbox{Nicolai~Popov}}
\altaffiliation{\texttt{science.popov@googlemail.com}}
\affiliation{Sektion Physik der Ludwig-Maximilians-Universit\"at, 4  Schellingstrasse, 80799 M\"unchen,
        Germany}

\author{\mbox{William~J.~Briscoe}}
\altaffiliation{\texttt{briscoe@gwu.edu}}
\affiliation{Institute for Nuclear Studies, Department of Physics, 
        The George Washington University, Washington, DC 20052, USA}

\author{\mbox{Igor~Strakovsky}}
\altaffiliation{\texttt{igor@gwu.edu}}
\affiliation{Institute for Nuclear Studies, Department of Physics, 
        The George Washington University, Washington, DC 20052, USA}

\noaffiliation

\begin{abstract}
The CP violation problem has a long history emanating from its discovery 60 years ago in the decay of neutral kaons and subsequent experimental and theoretical studies over several decades. We review herein experimental data that observe indirect CP violation of the order of $\sim10^{-3}$, as well as the discovery of direct CP violation of the order of $\sim10^{-6}$. Despite improved experimental methods over the past half a century, the original CP violation numbers have remained the same. Verification of the CP violation in the decay of charged kaons was also observed. Data reflecting CP violation in the decays of $B$ and $D$ mesons have become very important and are also discussed in this review. 
The question of CP violation only with the participation of $s\bar{s}$, $c\bar{c}$, and $b\bar{b}$ quarks in the framework of the Standard Model or beyond it and its small magnitude remains open.
\end{abstract}

\maketitle

\section{Introduction}
\label{Sec:Intro}
New particles, called K-mesons, or kaons, were discovered in cosmic rays in the late 1940s and were then studied more intensively at charged-particle accelerators beginning in 1954. These are the charged $K^\pm$ mesons (which form a particle / antiparticle pair) and the neutral $K^0$ meson (formed by a strange quark and a down antiquark);
with its antiparticle $\bar{K}^0$. Kaons have zero spin and a mass approximately 970 times that of an electron. Since they participate via strong interactions, they are identified as hadrons, but with zero baryon charge. They also are assigned a new quantum number, strangeness $S$, namely $S=1$ for $K^0$  and $S=-1$ for $\bar{K}^0$, characterizing their behavior in strong interaction processes. Until 1956, the law of conservation of parity was considered fundamental, as was the law of conservation of energy and momentum. However, Lee and Yang's suggestion of spatial parity non-conservation in weak interactions~\cite{Lee:1956qn} was confirmed in 1957 by Wu \textit{et al.} in studying the beta decay of Co$^{60}$ nuclei at NBS (now NIST)~\cite{Wu:1957my}. 

An important result of the study of $K^0$ mesons was the discovery of CP parity violation, or T parity violation, according to the CPT theorem~\cite{Luders:1954zz, Pauli:1955w}. Here, $P$ is a spatial inversion - parity, that is, a mirror image of the coordinates, replacing vector signs with the opposite ones for polar vectors $\vec{r}\to -\vec{r}$, as well as momentum $\vec{p}\to -\vec{p}$, while for pseudovectors, the angular momentum and spin do not change. $C$ is the charge conjugation parity, with the replacement of particles by their corresponding antiparticles, \textit{i.e.}, a change in the electric charge, hypercharge, baryon and lepton charges and, as a consequence, a change in the sign of the magnetic moment. In this case, the spatial coordinates, momentum, and spin of the particle do not change. $T$ denotes time reversal, $t\to -t$, that is, the reversibility of movement, with a corresponding change in the signs of momentum and spins. These parities are conserved in gravitational, electromagnetic and strong interactions but are violated in the weak interaction. The CTP theorem states that all quantum field theories must be symmetric under a combined transformation of P, C, and T which has been verified (see below).

\vspace{1cm}
\section{CP Parity}
\label{Sec:CP_Parity}
Neither $K^0$ nor $\bar{K}^0$ have a certain CP parity, but they do have negative spatial parity $P$. The action of the operators $C$, $P$, and CP on the wave functions $K^0$ and $\bar{K}^0$ can be written in the following form:
\begin{equation}
    C| K^0 \bigr \rangle = - | \bar{K}^0 \bigr \rangle \>,~~~~~ C| \bar{K}^0 \bigr \rangle = - | K^0 \bigr \rangle
     \>,
\nonumber
\end{equation}
\begin{equation}
    P| K^0 \bigr \rangle = - | K^0 \bigr \rangle \>,~~~~~ P| \bar{K}^0 \bigr \rangle = - | \bar{K}^0 \bigr \rangle
     \>,
\nonumber
\end{equation}
\begin{equation}
    CP| K^0 \bigr \rangle =  | \bar{K}^0 \bigr \rangle \>,~~~~~ CP| \bar{K}^0 \bigr \rangle =  | K^0 \bigr \rangle
     \>,
\nonumber
\end{equation}
where it is possible that antiparticles $K^0$ and $\bar{K}^0$ are transitioning from one to the other. Thus, from the states $K^0$ and $\bar{K}^0$ it is possible to construct linear combinations having definite CP parity values. These are
\begin{equation}
     | K_1^0 \bigr \rangle = \sqrt{\frac{1}{2}}~\bigl( | K^0 \bigr \rangle + | \bar{K}^0 \bigr \rangle \bigr) 
     \>,
\nonumber
\end{equation}
with CP parity $+1$, which can decay into two pions, and the orthogonal superposition
\begin{equation}
     | K_2^0 \bigr \rangle = \sqrt{\frac{1}{2}}~\bigl( | K^0 \bigr \rangle - | \bar{K}^0 \bigr \rangle \bigr),
\nonumber
\end{equation}
has CP parity $-1$, that can decay into two pions only as a result of a CP violating process such as the weak interaction. $K_1^0$ and $K_2^0$ are not particle -- antiparticle pairs and thus may have different decay channels. The resulting CP parities of $K_1^0$ and $K_2^0$ are:
\begin{equation}
    CP| K_1^0 \bigr \rangle = + | K_1^0 \bigr \rangle
    \>,
\nonumber
\end{equation}
\begin{equation}
    CP| K_2^0 \bigr \rangle = - | K_2^0 \bigr \rangle
    \>.
\nonumber
\end{equation}
Let us consider the conservation of combined CP-parity in the decays of $K$-mesons into 2 and 3 pions. 2$\pi$ and 3$\pi$ systems with orbital momentum $l = 0$ are eigenstates of the CP-operator. The operator P is equivalent to replacing $\pi^+\pi^-$  with the acquisition of the wave function of the multiplier $(-1)^l$.
\begin{equation}
    P| \pi^+\pi^- \bigr \rangle = P| \pi^+ \bigr \rangle ~P| \pi^- \bigr \rangle ~(-1)^l = + | \pi^+\pi^- \bigr \rangle
    \>.
\nonumber
\end{equation}
The operator $C$ turns $\pi^+$ into $\pi^-$ and $\pi^-$ into $\pi^+$, that is, it is also equivalent to replacement $\pi^+\pi^-$ mesons
\begin{equation}
    C| \pi^+\pi^- \bigr \rangle = (-1)^l~| \pi^+\pi^- \bigr \rangle = + | \pi^+\pi^- \bigr \rangle
\nonumber
\end{equation}
for $l=0$. And the eigenvalue of the CP operator of the 2$\pi$  system is equal to $+1$:
\begin{equation}
    CP| \pi^+\pi^- \bigr \rangle = (-1)^{2l}~| \pi^+\pi^- \bigr \rangle = + | \pi^+\pi^- \bigr \rangle
    \>.
\nonumber
\end{equation}
The CP parity of a system of three pion state depends on the orbital state $\vec{L}$ of the 3$\pi$ which is the sum of the orbital momentum vectors $\vec{l}$ of $\pi^+\pi^-$ pair and the orbital momentum $\vec{l}'$ of $\pi^0$ relative to the center of mass of this pion pair, that is, 
$\vec{L} = \vec{l} + \vec{l}' = 0$\ ($\vec{l}' = -\vec{l}$). 
\begin{equation}
    P| \pi^+\pi^-\pi^0 \bigr \rangle = P| \pi^+ \bigr \rangle ~P| \pi^- \bigr \rangle ~P| \pi^0 \bigr \rangle~(-1)^L = (-1) | \pi^+\pi^-\pi^0 \bigr \rangle
    \>,
\nonumber
\end{equation}
\begin{equation}
    C| \pi^+\pi^-\pi^0 \bigr \rangle = | \pi^+\pi^-\pi^0 \bigr \rangle~(-1)^L = (+1)| \pi^+\pi^-\pi^0 \bigr \rangle
    \>,
\nonumber
\end{equation}
\begin{equation}
    CP| \pi^+\pi^-\pi^0 \bigr \rangle = (-1)~| \pi^+\pi^-\pi^0 \bigr \rangle
    \>
\nonumber
\end{equation}
for $L=0$. In the result, the eigenvalue of the CP operator of the 3$\pi$ system is equal to $-1$. Since 2$\pi$ and 3$\pi$ systems are eigenvalues of the CP operator at zero orbital momentum, we assume in above relations that this momentum equals zero. 

The states $K_1^0$ and $K_2^0$ have fixed CP parity values, without, however, having fixed strangeness. We can write the following.
\begin{equation}
     | K^0 \bigr \rangle = \sqrt{\frac{1}{2}}~\bigl( | K_1^0 \bigr \rangle + | K_2^0 \bigr \rangle \bigr) 
     \>,
\nonumber
\end{equation}
\begin{equation}
     | \bar{K}^0 \bigr \rangle = \sqrt{\frac{1}{2}}~\bigl( | K_1^0 \bigr \rangle - | K_2^0 \bigr \rangle \bigr) 
     \>.
\nonumber
\end{equation}
That is $K^0$ and $\bar{K}^0$ are superpositions of the states $K_1^0$ and $K_2^0$. Since $CP(K_1^0) = +1$, while conserving the combined 
parity, with an average lifetime of the 2$\pi$ decay is $\tau(K_1^0) \approx 0.9\times 10^{-10}~\mathrm{s}$. Correspondingly, $CP(K_2^0) = -1$ and decays to 3$\pi$s with with a lifetime $\tau(K_2^0) \approx 5\times 10^{-8}~\mathrm{s}$ due to the smaller phase volume of the decay
products. Due to the low energy release in 3$\pi$ decays, their probabilities are approximately three orders of magnitude lower than the probabilities of 2$\pi$ decays. Since 50\% of the $K^0$ meson consists of the $K_1^0$ component, its decays into 2$\pi$ are observed near the target. At a greater distance from the target, decays of the $K_2^0$ component are observed in 3$\pi$ mesons. 

The inter-conversion of $K^0$ and $\bar{K}^0$ mesons in the vacuum is their oscillation between two superimposed states $K^0 \leftrightarrow \bar{K}^0$. Being a process with a $\Delta S = 2$ change in strangeness, ($S = 1$ for the $K^0$ meson and $S = -1$ for the $\bar{K}^0$ meson), this can only occur through weak interactions between the quarks from which the $K^0$ mesons are formed: $K^0 = \bar{s}d \leftrightarrow s\bar{d} = \bar{K}^0$.  As a result of these transformations, $K^0$ and $\bar{K}^0$ do not have fixed mass and lifetime. 
The allowed region in the $m_{K^0} - m_{\bar{K}^0}$ could  be obtained from uncertain of determination of $\delta$~\cite{ParticleDataGroup:2024pth}: 
\begin{equation}
    -4\times 10^{-19}~\mathrm{GeV} < m_{K^0} - m_{\bar{K}^0} < +4\times 10^{-19}~\mathrm{GeV}~(0.5\%~\mathrm{C.L.})
     \>.
\nonumber
\end{equation}
The kaon states with fixed mass and lifetime are short-lived $K_S^0$ and long-lived $K_L^0$ mesons. 

The difference between the masses of $K_L^0$ and $K_S^0$ is due to the weak interaction that causes $K^0 \leftrightarrow \bar{K}^0$ transitions; this mass difference is very small~\cite{ParticleDataGroup:2024pth}:
\begin{equation}
    \Delta m_K = m(K_L^0) - m(K_S^0) = (3.510\pm 0.018)\times 10^{-12}~\mathrm{MeV}
     \>.
\nonumber
\end{equation}
In the presence of CP invariance, we can consider: $| K_S^0 \bigr \rangle \equiv | K_1^0 \bigr \rangle$ and $| K_L^0 \bigr \rangle \equiv | K_2^0 \bigr \rangle$.

Let us assume that at an initial $t=0$ there exist pure $K^0$ mesons in the vacuum, \textit{e.g.}, produced in reaction $\pi^-p\to K^0\Lambda$ in the absence of $\bar{K}^0$ mesons. Then after time $t$ there will be the following beam:
\begin{equation}
    \frac{1}{2}~\Bigl[(K^0 + \bar{K}^0)~e^{-i(m_1 - \Gamma_1/2)t} + (K^0 - \bar{K}^0)~e^{-i(m_2 - \Gamma_2/2)t} \Bigr]
     \>.
\nonumber
\end{equation}
If $K_1^0$ and $K_2^0$ are stable, $\Gamma_1 = \Gamma_2 = 0$, the probabilities of finding  
\begin{equation}
    P_{K^0}(t) = \frac{1}{4}~|e^{-im_1t} + e^{-im_2t}|^2 = \cos^2 \Bigl( \frac{\Delta m_K}{2}t \Bigr)
     \>,
\nonumber
\end{equation}
\begin{equation}
    P_{\bar{K}^0}(t) = \frac{1}{4}~|e^{-im_1t} - e^{-im_2t}|^2 = \sin^2 \Bigl( \frac{\Delta m_K}{2}t \Bigr)
     \>
\nonumber
\end{equation}
and after time $\tau = \pi(m_2 - m_1) = \pi/\Delta m_K$  (oscillation period) there would be no $K^0$  mesons in beam, since:
\begin{equation}
    P_{K^0}(\tau) = \cos^2 \Bigl( \frac{\Delta m_K}{2}\tau \Bigr) = \cos^2 \frac{\pi}{2} = 0
     \>,
\nonumber
\end{equation}
\begin{equation}
    P_{\bar{K}^0}(\tau) = \sin^2 \Bigl( \frac{\Delta m_K}{2}\tau \Bigr) = \sin^2 \frac{\pi}{2} = 1
     \>.
\nonumber
\end{equation}

Such oscillations, which are superimposed by exponential decay due to $\Gamma_{1,2} \neq 0$, are observed experimentally.  It is possible to
observe them, for example, in semileptonic decays of neutral kaons that satisfy the rule $\Delta Q = \Delta S$. This rule is due to the fact that during the transition $s\to u$  the strangeness and electric charge of the quark increase by one. 
\begin{equation}
    K^0 \Bigl(\frac{d}{\bar{s}} \Bigr) \to \pi^- \Bigl(\frac{d}{\bar{u}} \Bigr)~l^+\nu_l \>,~~~ 
    K^0 \Bigl(\frac{d}{\bar{s}} \Bigr) \nrightarrow \pi^+ \Bigl(\frac{u}{\bar{d}} \Bigr)~l^-\bar{\nu}_l
     \>,
\nonumber
\end{equation}
\begin{equation}
    \bar{K}^0 \Bigl(\frac{s}{\bar{d}} \Bigr) \to \pi^+ \Bigl(\frac{u}{\bar{d}} \Bigr)~l^-\bar{\nu}_l  \>,~~~ 
    \bar{K}^0 \Bigl(\frac{s}{\bar{d}} \Bigr) \nrightarrow \pi^- \Bigl(\frac{d}{\bar{u}} \Bigr)~l^+\nu_l
     \>.
\nonumber
\end{equation}
Here $l  \equiv e,\ \mu$. The sign of the charged lepton indicates the strangeness of $K^0$ or $\bar{K}^0$. Oscillations can be measured by 
recording the numbers of electrons and positrons arising from $K_{e3}$ decays; such experiments make it possible to determine $\Delta m_K$. Reactions with $\nrightarrow$ are prohibited by the rule $\Delta Q = \Delta S$  and have not been observed experimentally.

The $K^0$-meson, arising as a result of the strong interaction, at some distance from the initial production point is practically transformed due to the weak interaction into $\bar{K}^0$ and therefore turns out to be capable of causing nuclear reactions, for example, $\bar{K}^0p\to \Lambda \pi^+$ characteristic of $\bar{K}^0$ and prohibited for $K^0$ due to the preservation of strangeness in strong interaction. In 1955, Pais and Piccioni predicted oscillations in a beam of neutral $K$ mesons within the framework of the regeneration effect~\cite{Pais:1955sm}. Regeneration consisted of the use of an absorber, positioned at a distance determined by $\tau_S$ and $\tau_L$ to demonstrate the superposition of $K^0$ and $\bar{K}^0$. Let us take a pure beam $K^0$ in vacuum (equal of parts $K_S^0$ and $K_L^0$). After, \textit{e.g.}, $t\approx 10\ \tau_S$, the intensity of $K_S^0$ was reduced by a factor of $e^{-t/\tau_S} = e^{-10} \approx 4.5\times 10^{-5}$ ($0.5~\mathrm{m}$ for the $1~\mathrm{GeV}$ kaons).
The $6.2~\mathrm{GeV}$ proton synchrotron at Lawrence Berkeley National Laboratory (LBNL) with a beam $\pi^-$ of $1.1~\mathrm{GeV}$ was used for the experiment.

A beam of neutral K-mesons decays in flight so that the short-lived $K_S^0$ disappears, leaving a long-lived pure beam $K_L^0$. If this beam passes through matter, $K^0$ and $\bar{K}^0$ interact with nuclei in different ways. With $K^0$, quasi-elastic scattering on nucleons occurs, while $\bar{K}^0$ can produce hyperons. Due to the different interactions of these components, the quantum coherence between the two particles is lost. The resulting beam contains various linear superpositions of $K^0$  and $\bar{K}^0$. Each superposition is a mixture of $K_S^0$ and $K_L^0$, so $K_S^0$ is restored when a beam of K-mesons passes through the substance of the regenerator, as was observed in the above experiment. Anomalous regeneration of $K_S^0$  was obtained in Ref.~\cite{Leipuner:1963zz}.

The problem of time reversal in quantum mechanics has been discussed since the 1930s~\cite{Wigner:1932e}. In
1964, Christenson, Cronin, Fitch, and Turlay discovered CP violation in an experiment at the Brookhaven National Laboratory (BNL) AGS accelerator~\cite{Christenson:1964fg} through the decay of the neutral $K_L^0$  
meson to 2$\pi$  mesons in spite of the main decay channel to 3$\pi$. In fact, there are two K-mesons: $K_1^0\to  \pi^+\pi^-$ with $CP = +1$ (while $CP = +1$ for the state), that is, the even state, and $K_2^0\to \pi^+\pi^-\pi^0$ with $CP = -1$, \textit{i.e.}, the odd state. 

The long-lived $K_L^0$  has a probability of approximately 0.2\% decaying into 2$\pi$ with CP parity violation, which is small, but is, however, an important discovery for fundamental physics~\cite{ParticleDataGroup:2024pth}:
\begin{equation}
    \frac{\Gamma(K_L^0\to \pi^+\pi^-)}{\Gamma(K_L^0\to all)} = (2.03\pm 0.04)\times 10^{-3}
     \>.
\nonumber
\end{equation}

\vspace{0.3cm}
CP violation in mixtures of neutral mesons $K^0 \leftrightarrow \bar{K}^0$ is an indirect violation of CP parity, in contrast to direct violation at different decay rates of a particles and antiparticles and interference between direct and indirect violation of CP parity. There is no strict identification of $K_S^0$  with $K_1^0$ and $K_L^0$ with $K_2^0$. Instead, the states $K_L^0$ and $K_S^0$  are defined as follows:
\begin{equation}
     | K_L^0 \bigr \rangle = \Bigl(1 + |\epsilon|^2 \Bigr )^{-1/2}~\Bigl(| K_2^0 \bigr \rangle 
     + \epsilon | K_1^0 \bigr \rangle \Bigr ) 
     = \Bigl[ 2 \Bigl(1 + |\epsilon |^2 \Bigr ) \Bigr ]^{-1/2} \Bigl[(1 
     + \epsilon)| K^0 \bigr \rangle - (1 - \epsilon) | \bar{K}^0 \bigr \rangle \Bigr ]
     \>,
\nonumber
\end{equation}
\begin{equation}
     | K_S^0 \bigr \rangle = \Bigl(1 + |\epsilon|^2 \Bigr )^{-1/2}~\Bigl(| K_1^0 \bigr \rangle 
     + \epsilon | K_2^0 \bigr \rangle \Bigr ) 
     = \Bigl[ 2 \Bigl( 1 + |\epsilon |^2 \Bigr ) \Bigr ]^{-1/2} ~\Bigl[(1 
     + \epsilon)| K^0 \bigr \rangle + (1 - \epsilon) | \bar{K}^0 \bigr \rangle  \Bigr ]
     \>.
\nonumber
\end{equation}

The laws of nature would be the same for matter and for antimatter. Most phenomena are $C$- and $P$-symmetric, and also CP-symmetric for electromagnetic, strong, and gravitational interactions. The weak interactions violate the $C$ - and $P$-symmetric as well as some CP-symmetry of unknown nature with the supposition of a superweak interaction. At the same time, the Universe is made chiefly of matter, rather than consisting of equal parts of matter and antimatter, as might be expected. 

\section{Matter-Antimatter Problem}
\label{Sec:M-AM}
The Universe is made chiefly of matter, rather than consisting of equal parts of matter and antimatter, as might be expected. In order to create an imbalance in matter and antimatter from an initial balance, the special conditions must be satisfied.

In 1967, Sakharov showed that CP violation was one of the necessary conditions for the almost complete destruction of antimatter at the early stage of the development of the Universe in the first seconds following the Big Bang~\cite{Sakharov:1967dj, Popov:2021n}. The Big Bang should have
produced equal amounts of matter and antimatter that would have resulted in the Universe being filled with radiation due to complete matter-antimatter annihilation. Our Universe is filled with a large number of \textit{relic photons} formed as a result of partial matter-antimatter annihilation as is recorded by radio telescopes as the observed Cosmic Microwave Background, the 2.72548±0.00057~K radiation that fills the entire Universe. Obliviously, not all of the matter was annihilated into photons; about one of every billion quarks survived and is the origin of the Universe that exists today. This leads to some questions: how could some matter survive the primordial annihilation; why did matter survive instead of antimatter;  and where and when did the corresponding antimatter disappear?  Sakharov’s hypothesis is that after the Big Bang, the physical laws acted differently for matter and antimatter, that is, the CP symmetry was violated. 

It can be shown that to create the current imbalance between matter and antimatter, the Sakharov conditions must be satisfied; that is, violation of the baryon number, violation of symmetry $C$, violation of CP symmetry, and the interaction process is outside of that allowed by thermal equilibrium. Baryon number violation is a necessary condition for producing an excess of baryons over antibaryons. One of its conditions is the violation of CP symmetry in the extreme conditions of the first seconds after the Big Bang. Since our Universe is not a sea of photons without other matter, immediately after the Big Bang different physical laws than now exist now were in effect for matter and antimatter: that is, for one, CP symmetry was violated. However, breaking the $C$ symmetry was also necessary so that interactions that produced
more baryons than antibaryons were not balanced by interactions that produced more antibaryons than baryons. Breaking of CP symmetry was necessary -- otherwise an equal number of left baryons and right
antibaryons would be produced, as well as an equal number of right baryons and left antibaryons.  These interactions had to be outside of the limitations of thermal equilibrium; otherwise, CPT symmetry would have compensated for the processes of decreasing or increasing the baryon number. 

\section{CP Violation in Standard Model}
\label{Sec:SM}
In 1973, in their attempt to find an explanation for the violation of the CP in the decay of neutral kaons using the idea of Cabibbo concerning the mixing of two generations of quarks, Kobayashi and Maskawa predicted the existence of a third generation of quarks. As a result of their work, the bottom $b$ quark was discovered in 1977~\cite{E288:1977xhf}, and the $t$ quark in 1995~\cite{D0:1995jca, CDF:1995wbb}.

\begin{figure*}[htb!]
\vspace{-0.3cm}
\centering
{
    \includegraphics[width=0.6\textwidth,keepaspectratio]{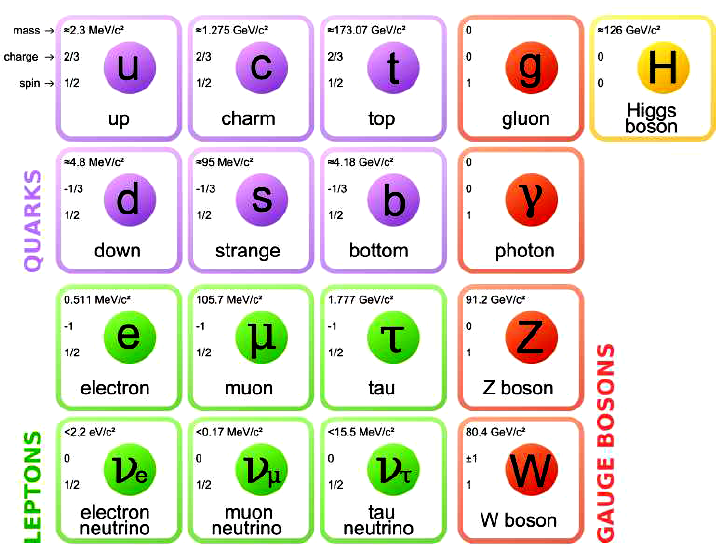} 
}

\centerline{\parbox{0.8\textwidth}{
\caption[] {\protect\small
The Standard Model of elementary particles, taken from Ref.~\cite{ParticleDataGroup:2024pth}.
} 
\label{fig:fig1} } }
\end{figure*}

Possible CP violation is associated with complex weak coupling constants because of the CP-odd interaction of
participating particles. The CP violation appears naturally in united gauge theories with spontaneous symmetry
breaking.  This mechanism is possible through the complex vacuum expectation value of the Higgs field \textcolor{red}{presented} in the Standard Model (SM) (Fig.~\ref{fig:fig1}). The simplest model of CP breaking through the Higgs field coupling results in a complex phase in the Cabibbo-Kobayashi-Maskawa mass matrix (CKM)~\cite{Cabibbo:1963yz, Kobayashi:1973fv}.

The interaction of quarks in the SM with charged gauge bosons $W^\pm$ is described by:
\begin{equation}
    L = - \frac{g}{2\sqrt{2}}~\bar{u}_i~\gamma^\mu~(1 - \gamma_5)~V_{ij}~d_j~W_\mu^\pm~ + h.c.
     \>,
\nonumber
\end{equation}
where $g$ is the weak interaction constant, $u_i$ and $d_j$ is the fields of quark states of three generations,   $W^\pm$ is the fields of vector charged bosons, and $V_{ij}$ is the quark mixing matrix, that is, the CKM matrix:
\begin{equation}
    V = \left( \begin{array}{ccc} c_{12}c_{13}    & s_{12}c_{13}                                 & s_{13}e^{-i\delta}\\
    -s_{12}c_{23} - c_{12}s_{23}s_{13}e^{i\delta} & c_{12}c_{23} - s_{12}s_{23}s_{13}e^{i\delta} & s_{23}c_{13}\\
     s_{12}s_{23} - c_{12}c_{23}s_{13}e^{i\delta} &-c_{12}s_{23} - s_{12}c_{23}s_{13}e^{i\delta} & c_{23}c_{13}
     \end{array} \right)
    \>.
\nonumber
\end{equation}
Here $c_{ij} = \cos\theta_{ij}$, $s_{ij} = \sin\theta_{ij}$ and a single phase $\delta$ in the interval [0, 2$\pi$]~\cite{ParticleDataGroup:2024pth}. The CKM matrix uses three Euler angles $\theta_{12}$, $\theta_{13}$, and $\theta_{23}$ and a CP violating phase $\delta$ with values:

$$\theta_{12} = (~13.04\pm 0.05~)^\circ\>,$$ 
$$\theta_{23} = (~~2.38\pm 0.06~)^\circ\>,$$ 
$$\theta_{13} = (0.201\pm 0.011)^\circ\>,$$
$$\delta      = (0.201\pm 0.011)^\circ\>.$$

In the case of two generations of quarks, the mixing matrix is parameterized by a single Cabibbo angle $\theta_{12} = \theta_c = 13.02^\circ$:
\begin{equation}
    V = \left( \begin{array}{cc} \cos\theta_c & \sin\theta_c \\
                                -\sin\theta_c & \cos\theta_c \end{array} \right)
    \>,
\nonumber
\end{equation}
there is no possible mechanism for CP violation. The presence of a third generation of quarks, for example, the $t$ quark, would be a necessary condition for violation of the CP parity.

The standard CKM matrix parametrization given above is derived from the original CKM:
\begin{equation}
    V = \left( \begin{array}{ccc} V_{ud} & V_{us} & V_{ub}\\
                                  V_{cd} & V_{cs} & V_{cb}\\
                                  v_{tb} & V_{ts} & V_{tb}
     \end{array} \right)
    \>
\nonumber
\end{equation}
reduced to the four indicated parameters determined from the experiment.  The matrix for the generation of $N$ quarks, or flavors $2N$, contains $N^2$ complex numbers or 2$N^2$ real parameters. In SM of three-quark generation $3\times 3$ matrix $V$ contains 9 complex numbers or 18 real parameters. The unitarity condition:
\begin{equation}
    VV^\ast = 1
     \>,
\nonumber
\end{equation}
for the $N\times N$-matrix (where $V^\ast$ is the conjugate transpose of $V$) requires $N^2$ real parameters to be specified. Indeed, from 
\begin{equation}
    \sum_k V_{ik}V^\ast_{jk} = \delta_{ij}
     \>
\nonumber
\end{equation}
for the diagonal components with $i = j$ there are $N$ restrictions and $N(N - 1)$ for $i\neq j$, \textit{i.e.}, outside the diagonals ($\frac{1}{2}N(N - 1)$ for $Re(V_{ik}V^\ast_{jk})$ and $Im(V_{ik}V^\ast_{jk})$). So, the number of parameters remaining is $N + N(N - 1) = N^2$. From the $N^2$, the remaining parameters $2N - 1$, are not physically significant, because one phase can be absorbed into each quark field (\textit{i.e.}, the mass and weak eigenstates), but the matrix is independent of a common phase. So, the total number of free parameters, independent of the choice of the phases of base vectors, is
\begin{equation}
    N^2 - (2N - 1) = (N - 1)^2
     \>.
\nonumber
\end{equation}
Of these, $\frac{1}{2}N(N - 1)$ are rotation angles, \textit{i.e.}, quark mixing angles. The remaining $\frac{1}{2}(N - 1)(N - 2)$ are the complex phases that cause CP violation.  In the SM case ($N = 3$), three mixing Euler angles are indicated above ($\theta_{12}$, $\theta_{13}$ and $\theta_{23}$) and one complex phase CP violating $\delta$.

The CKM matrix describes the probability of transition from flavour quark to another one, $|V_{ij}|^2$,   with experimental elements~\cite{ParticleDataGroup:2024pth}:
\begin{equation}
    \left( \begin{array}{ccc} |V_{ud}| & |V_{us}| & |V_{ub}|\\
                              |V_{cd}| & |V_{cs}| & |V_{cb}|\\
                              |v_{tb}| & |V_{ts}| & |V_{tb}|
     \end{array} \right)
  = \left( \begin{array}{ccc} 0.97435\pm 0.00016 & 0.22501\pm 0.00068 & 0.003732_{-0.000085}^{+0.000090}\\
                              0.22187\pm 0.00068 & 0.97349\pm 0.00016 & 0.04183_{-0.00069}^{+0.00079}\\
                              0.00858_{-0.00017}^{+0.00019} & 0.04111_{-0.00068}^{+0.00077} & 0.999118_{-0.000034}^{+0.000029}
     \end{array} \right)
    \>.
\nonumber
\end{equation}

The verification of unitarity of CKM matrix indicates:   
\begin{equation}
    |V_{ud}|^2 + |V_{us}|^2 + |V_{ub}|^2 = 0.9985\pm 0.0007
     \>
\nonumber
\end{equation}
in excellent agreement with unitarity.

It is known experimentally that $s_{13} \ll s_{23} \ll s_{12} \ll 1$, and it is convenient to present this hierarchy using the Wolfenstein parametrization~\cite{Wolfenstein:1983yz}:
\begin{equation}
    \lambda = s_{12} = \frac{|V_{us}|}{\sqrt{|V_{ud}|^2 + |V_{us}|^2}}
     \>,
\nonumber
\end{equation}
\begin{equation}
    A\lambda^2 = s_{23} = \lambda\Big|\frac{V_{cb}}{V_{us}}\Big|
     \>.
\nonumber
\end{equation}
\begin{equation}
    A\lambda^3(p + i\eta) = s_{13}e^{i\delta} = V^\ast_{ub}
     \>.
\nonumber
\end{equation}
The Wolfenstein parametrization corresponds to an expansion in $\lambda$:
\begin{equation}
    V = \left( \begin{array}{ccc} 1 - \frac{1}{2}\lambda^2    & \lambda                   & A\lambda^3(\rho - i\eta)\\
                                        -\lambda              & 1 - \frac{1}{2}\lambda^2  & A\lambda^2\\
                                  A\lambda^3(1 - \rho -i\eta) & -A\lambda^2               & 1
     \end{array} \right)
    + O(\lambda^4)
    \>.
\nonumber
\end{equation}
This approximation on the order of $\lambda^3$ has a better than 0.3\% accuracy. The CP violation rates correspond to the parameters $\rho$ and $\eta$. Wolfenstein parameters are the following~\cite{ParticleDataGroup:2024pth}:
$$\lambda = 0.22501\pm 0.00068 \>,$$
$$A = 0.826^{+0.016}_{-0.015} \>,$$
$$\bar{\rho} = 0.1591\pm 0.0094 \>,$$
$$\bar{\eta} = 0.3523^{+0.0073}_{-0.0071} \>.$$

The unitarity constraint on the CKM matrix can be written as $\sum_k V_{ik}V_{jk}$ for $i\neq j$ is a constraint on three complex numbers, one for each $k$, which form the sides of a triangle in the complex plane. There are six choices of $i$ and $j$, three of which are independent, and six unitary triangles. The deviation areas of these triangles from equality determine the degree of CP violation. The most commonly used unitarity triangle arises from 
\begin{equation}
    V_{ud}V^\ast_{ub} + V_{cd}V^\ast_{cb} + V_{td}V^\ast_{tb} = 0
     \>.
\nonumber
\end{equation}
If we assume that real parts of $V_{ud}$, $V_{cb}$, and $V_{tb} > 0$ and real part of $V_{cd} < 0$ (see Wolfenstein matrix for $-\lambda$). So:
\begin{equation}
    |V_{ud}||V^\ast_{ub}| - |V_{cd}||V_{cb}| + |V_{td}||V_{tb}| = 0
     \>.
\nonumber
\end{equation}
\begin{equation}
    1 - \bigg|\frac{V_{ud}V_{ub}}{V_{cd}V_{cb}}\bigg| - \bigg|\frac{V_{td}V_{tb}}{V_{cd}V_{cb}}\bigg| = 0 
     \>.
\nonumber
\end{equation}

\begin{figure*}[htb!]
\vspace{-0.3cm}
\centering
{
    \includegraphics[width=0.5\textwidth,keepaspectratio]{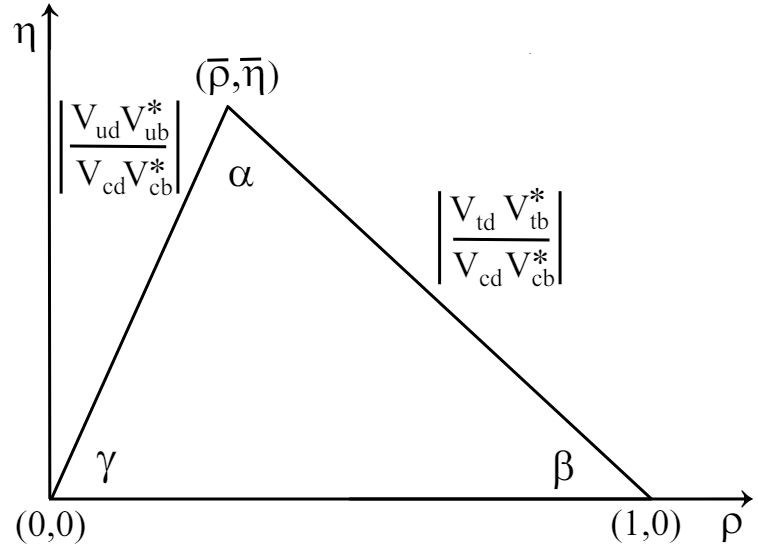} 
}

\centerline{\parbox{0.8\textwidth}{
\caption[] {\protect\small
        The unitarity triangle. 
} 
\label{fig:fig2} } }
\end{figure*}

We can put this relation in the complex plane and interpret it as a unitarity triangle (see Fig.~\ref{fig:fig2}). 
Here,
\begin{equation}
    \bar{\rho} = \rho(1 - \lambda^2/2 + ...)
     \>,
\nonumber
\end{equation}
\begin{equation}
    \bar{\eta} = \eta(1 - \lambda^2/2 + ...)
     \>,
\nonumber
\end{equation}
The angles are
\begin{equation}
    \alpha = arg\Big(-\frac{V_{td}V^\ast_{tb}}{V_{ud}V^\ast_{ub}}\Big) = (84.1^{+4.5}_{-3.8} )^\circ
     \>,
\nonumber
\end{equation}
\begin{equation}
    \beta = arg\Big(-\frac{V_{cd}V^\ast_{cb}}{V_{td}V^\ast_{tb}}\Big) = (22.6^{+0.5}_{-0.1})^\circ
     \>,
\nonumber
\end{equation}
\begin{equation}
    \gamma = arg\Big(-\frac{V_{ud}V^\ast_{ub}}{V_{cd}V^\ast_{cb}}\Big) = (65.7\pm 3.0)^\circ
     \>.
\nonumber
\end{equation}
The amplitude of the CP violating asymmetry is proportional to $\sin2\beta = 0.709\pm 0.011$~\cite{ParticleDataGroup:2024pth}.

Within the framework of the SM using the CKM matrix, calculations of CP violating effects were carried out in the kaon system, as well as in systems with $B$ and $D$ mesons, where these effects are greater and were also discovered experimentally. Violation of the CP symmetry in decays of $B$ mesons was observed in 2001 in the BaBar experiment at SLAC and in experiments at the KEK facility. 

As for the baryon number, there is no experimental evidence of the violation of this number. The violation of such conservation in the SM is possible outside the framework of perturbation theory. Baryon number breaking is possible within the framework of supersymmetry or Grand Unification Theory (GUT).  

\section{Discovery of Indirect CP Violation}
\label{Sec:CP-indir}
\begin{figure*}[htb!]
\vspace{-0.3cm}
\centering
{
    \includegraphics[width=0.7\textwidth,keepaspectratio]{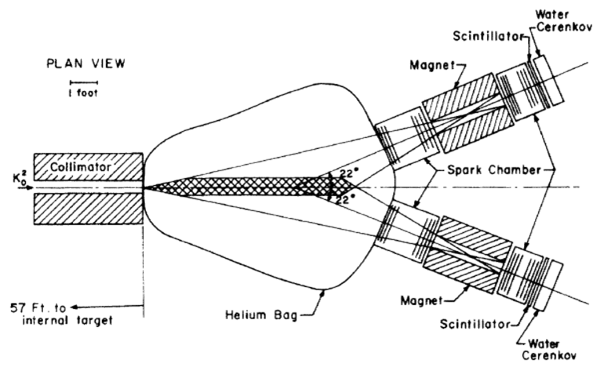} 
}

\centerline{\parbox{0.8\textwidth}{
\caption[] {\protect\small
Plan view of the detector arrangement, taken from Ref.~\cite{Christenson:1964fg}.
} 
\label{fig:fig3} } }
\end{figure*}

Figure~\ref{fig:fig3}, taken from Ref.~\cite{Christenson:1964fg}, presents a general overview of the
detector system of the experiment carried out at Brookhaven National Laboratory.  Spark chambers were located in each arm of
a two-arm spectrometer; these record the trajectories of particles before and after the deflecting magnet. Cherenkov
scintillation counters triggered the spark chambers when a coincidence was detected, and the resulting tracks were
photographed and recorded on film.  The detector system was installed in the beam of the $K^0$ meson beam line at the
Brookhaven synchrotron at a distance such that the $K_1^0$  mesons had already decayed so that only the $K_2^0$ mesons
remained in the beam. The angle between the arms was chosen for optimal determination of the kaons that decay into two
charged pions. Helium gas filled the volume where this decay occurred to minimize the detection of three-particle decays
of $K_2^0$ mesons by rescattering. In this case, the vector sum of the momenta of the two detected particles must be
oriented in the direction of the beam of $K_2^0$ mesons, and the missing mass of the decay particle, calculated from the
data on the decay products, must coincide with the mass of the K-meson. 56 desired events were recorded with a background
of 11 events.  From these data it follows that the fraction of the decay of $K_2^0$ into two pions in relation to all
decay modes involving charged particles is equal to $2\times 10^{-3}$. This was the first evidence of the violation of parity in the CP system in the kaon system. Almost six months were spent searching for an alternative CP violation, but to no avail. Thus, 
their conclusion that they obtained indirect CP violation (Fig.~\ref{fig:fig3}) in the mixing of neutral kaon states ($K^0 \leftrightarrow \bar{K}^0$).

\section{Direct and Indirect CP Violation}
\label{Sec:CP-dir}
CP violation for quarks with a charge of $-1/3$ corresponds to the strange $s$ quark that is a constituent of kaons and the beauty $b$ quark that is a constituent of $B$-mesons. However, CP violation was also predicted for quarks with charge of $+2/3$, corresponding to the charm c-quark that is a constituent of the $D$-meson. The weak force appears to act not on a quark state, as identified by the flavor of the quark, but on a quantum mixture of two types of quark for indirect CP violation in contrast to direct CP violation in the pure quark state.

In agreement with the tenants of Wolfenstein superposition,~\cite{Wolfenstein:1964ks} which was first postulated in 1964, any of several superweak interactions are possible. A direct consequence of this manifests itself only in the mixing $\Delta S =   2$ between $K^0$ and $\bar{K}^0$ with the absence in $\Delta S = 1$. A consequence of this assumption could be the possibility of correlations between the decay amplitudes of $K^0$. Let
\begin{equation}
    \eta^{+-} = \frac{A(K_L^0\to \pi^+\pi^-)}{A(K_S^0\to \pi^+\pi^-)} \>,~~~
    \eta^{00} = \frac{A(K_L^0\to \pi^0\pi^0)}{A(K_S^0\to \pi^0\pi^0)}
    \>.
\nonumber
\end{equation}

In this case, the decay amplitude of the CP-odd combination $K_2^0$ combination to a  2$\pi$ final state might differ from zero, corresponding to a direct violation of CP $\epsilon$. However, $\eta^{+-} = \eta^{00} = \epsilon$  in the case of indirect CP
violation. There are two decay amplitudes, one is the $I = 0$ and the second is found among the $I = 2$ states and final states. As the Clebsch-Gordan coefficients projecting, the $I = 0$ and $I = 2$ states for the $\pi^+\pi^-$ and 2$\pi^0$ final states are different, we have:    
\begin{equation}
    \epsilon' = i\sqrt{1/2}~Im\Bigl(\frac{A_2}{A_0} \Bigr)~e^{i(\delta_2 - \delta_0)}
    \>,
\nonumber
\end{equation}
here $A_{0,2}$ and $\delta_{0,2}$ are the amplitudes and the strong phases of the 2$\pi$ final isospin states $I = 0, 2$. Then, there exists direct CP violation, determined by complex parameter $\epsilon'$, which states that there is possible $K_2^0\to 2\pi$ in the absence above equality, \textit{i.e.}:
\begin{equation}
    \eta^{+-} = \epsilon + \epsilon' \>,~~~ \eta^{00} = \epsilon - 2\epsilon'
    \>,
\nonumber
\end{equation}

Modern experimental data, taken from Refs.~\cite{ParticleDataGroup:2024pth}, are the following:

$$\epsilon \equiv |\epsilon| = (2.228\pm 0.011)\times 10^{-3}\>,$$
$$Re(\epsilon)  = (1.66\pm 0.02)\times 10^{-3}\>,$$
$$Im(\epsilon)  = 1.57 \>,$$
$$Re(\epsilon') = (2.5\pm 0.4)\times 10^{-6}\>,$$
$$|\eta^{00}|   = (2.220\pm 0.011)\times 10^{-3}\>,$$
$$|\eta^{+-}|   = (2.232\pm 0.011)\times 10^{-3}\>,$$
$$|\eta^{00}/\eta^{+-}| = 0.9950\pm 0.0007\>,$$
$$\epsilon + \epsilon' = (2.232\pm 0.011)\times 10^{-3}\>,$$
$$\epsilon - 2\epsilon' = (2.221\pm 0.011)\times 10^{-3}\>,$$
$$Re(\epsilon'/\epsilon) = (1.65\pm 0.26)\times 10^{-3}\>,$$
$$\phi_{+-} = (43.51\pm 0.05)^\circ\>,$$
$$\phi_{00} = (43.52\pm 0.05)^\circ\>.$$
These phases are for $\eta^{+-} = |\eta^{+-}|e^{i\phi_{+-}}$ and $\eta^{00} = 
|\eta^{00}|e^{i\phi_{00}}$, summing CPT invariance~\cite{ParticleDataGroup:2024pth}. An estimate of the phases $\phi_{+-}$ and $\phi_{00}$ was given
by Bell and Steinberger in 1965~\cite{Bell:1966vvu} indicating their equality to the phase $\phi_\epsilon$ and taking into account the CPT theorem with the above-mentioned $\Delta m_K$ and $\Gamma_S = 1.11\times 10^{10}~\mathrm{s^{-1}} $ they are equal $\phi_\epsilon = 
43.5^\circ$ in agreement with the experimental values mentioned above. Experimental confirmation of the approximate equality $\eta_{+-} \approx \eta_{00}$ was obtained in the 1970s~\cite{Holder:1972zv, Banner:1972hx, Christenson:1979tt}. 
\begin{equation}
    Re\Bigl( \epsilon'/\epsilon \Bigr) = \frac{1}{6}\Bigl( 1 - \Big|\frac{\eta^{00}}{\eta^{+-}}\Big|^2 \Bigr) = \frac{1}{6} (1 - R)
    \>,
\nonumber
\end{equation}
where $R$ is the so-called double ratio:
\begin{equation}
    R = \frac{\Gamma(K_L^0 \to \pi^0\pi^0)}{\Gamma(K_S^0 \to \pi^0\pi^0)}  / 
        \frac{\Gamma(K_L^0 \to \pi^+\pi^-)}{\Gamma(K_S^0 \to \pi^+\pi^-)}
    \>.
\nonumber
\end{equation}

\begin{figure*}[htb!]
\vspace{-0.3cm}
\centering
{
    \includegraphics[width=0.6\textwidth,keepaspectratio]{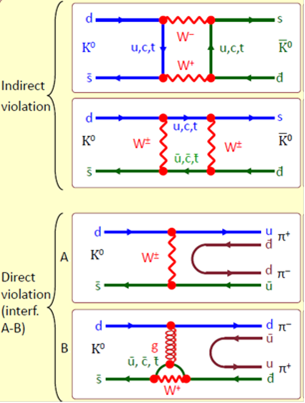} 
}

\centerline{\parbox{0.8\textwidth}{
\caption[] {\protect\small
The indirect and direct modes of CP violation.
} 
\label{fig:fig4} } }
\end{figure*}

After the discovery of indirect CP violation~\cite{Christenson:1964fg}, further experiments were carried out to find direct CP violation (see Fig.~\ref{fig:fig4}). There are two possible manners of violation of the CP: in the first, the long-lived kaon is a mixture of two quantum states, mainly CP-odd with just a small amount of CP-even; in the second, CP violation in decay of $K^0$ when the actual quark reactions underlying the particle transformations. In the decay of the neutral kaon, a strange quark disappears with creation of pions composed of only up- and down-quarks -- CP violation via this route is called a direct one. 
\vspace{0.5cm}

Direct CP violation necessitates that the probabilities of the decays of particles and antiparticles do not coincide; indirect CP violation in the processes of mixing of neutral mesons means that the probabilities of the direct and reverse processes are unequal. Additionally, interference is also possible when mixing the direct and indirect mechanisms of CP violation. While direct CP violation does not depend on the time from meson formation, the other two mechanisms are time-dependent. If the asymmetry changes with time, for example, in the decays $B^0\to 2\pi$ or $B^0\to 2K$, then the CP violation depends on time.

\section{CERN and Fermilab Experiments}
\label{Sec:CF}
In the NA31 experiment performed at CERN~\cite{NA31:1988eyf, NA31:1993tha}, the decay channels of the neutral and
charged kaons were simultaneously recorded both near the dipping volume and far from it ($K_L^0$). This made it
possible to avoid the background from the interaction of neutrons in the regenerator and to minimize the difference
in the geometries of the decay lengths of which differ by approximately three orders of magnitude. The energies of
neutral and charged particles were measured by two calorimeters filled with liquid argon. A detailed description of
the detectors in this experiment is given in Refs.~\cite{NA31:1988eyf, NA31:1993tha}. The short and long-lived 
kaons were born in collisions of protons of the $450~\mathrm{GeV}$ Super Proton Synchrotron (SPS) with two targets, 
one of which ($K_L^0$ target) was located at a large distance (about $100~\mathrm{m}$) from the decay volume, and 
the other  ($K_S^0$  target) close to it. The trigger for launching the selection process to isolate decays into
charged pions was triggered on the basis of data from the scintillation hodoscope located between the second wire
chamber and the electromagnetic calorimeter. The trigger for isolating decays into neutral pions was launching by a
hodoscope placed in liquid argon. The main problems of this experiment were related to the lack of a spectrometer
to measure charged tracks. In this case, about 40\% of the decays to 2$\pi$ were lost.

A characteristic of another experiment E731 (Fermilab)~\cite{E731:1993niw, E731:1997bzb} was the simultaneous use of two almost parallel $K_L^0$ beams, one of which, before the decay volume, passed through a B$_4$C  regenerator, where (due to the difference in the scattering of the nuclei $K^0$ and $\bar{K}^0$) coherent regeneration of the $K_L^0$ component occurred. A magnetic spectrometer was used to record the tracks of charged particles. Neutral decays were recorded using an electromagnetic calorimeter containing 804 glass cells. Photomultipliers recorded Cherenkov light from electrons and positrons generated in glass by photons or high-energy electrons. The trigger of charged decays was launched as a result of coincidences of signals in two opposite quadrants of the scintillation hodoscope. Compared with the NA31 experiment, it is clear that in the E731 experiment, the background from $K_L^0\to 3\pi^0$ in the neutral decay mode is significantly lower due to the closer location of the decay volume to the detectors.  There is also less background in charged mode from a detector with better resolution to measure the momentum of charged particles. At the same time, the result of the E731 experiment strongly depended on the correct accounting of the efficiency of decay detection performed by Monte Carlo simulation.  The results of these experiments are the following. NA31 (CERN) $Re(\epsilon'/\epsilon) = (2.30\pm 0.65)\times 10^{-3}$ and E731 $Re(\epsilon'/\epsilon) = (0.74\pm 0.61)\times 10^{-3}$. 
From a comparison of these experiments it is clear that they do not agree with each other (with a confidence level of less than 10\%). Experiment NA31 indicates the possibility of direct CP violation $\approx 10^{-3}$ , while the result of experiment E731 is compatible with zero.  Therefore, there was no clear conclusion about a direct violation of the CP invariance at that time, and further more accurate measurements were required.

The theoretical calculations obtained for the $\epsilon'/\epsilon$ ratio up to the 1980s ranged from 0.002 to 0.02~\cite{Ellis:1976fn, Gilman:1978wm, Guberina:1979ix}. Theoretical consideration of this problem continued in the 1990s with a decrease in the calculated $\epsilon'/\epsilon$ to $10^{-4}$~\cite{Buchalla:1989we, Ciuchini:1995cd, Buras:1996dq} with an indication of the reduction of the so-called loop ``penguin'' diagrams~\cite{Flynn:1989iu}. However, these diagrams are highly sensitive to possible contributions from unknown particles and therefore provide excellent probes for new sources of CP violation. The known source of CP violation, which is governed by the CKM matrix in the quark sector, is insufficient to account for the huge excess of matter over antimatter in the Universe; new sources of CP violation are necessary to solve this problem.

The NA48 experiment was significantly improved compared to the previous NA31 experiment with the participation of new institutions in the collaboration. The goal was to improve the precision of $\epsilon'/\epsilon$ to $10^{-4}$. Significant changes were made to the beam system and the detector. The NA48 experiment at CERN~\cite{NA48:2002tmj, NA48:2001bct} had made separate runs with long- and short-lived kaons, $K_L^0$ and $K_S^0$, produced on two different targets by protons from the same CERN SPS beam. The four modes of decay in the $K_L^0$ and $K_S^0$ beams reduced the systematic uncertainties in the measurement of decay probabilities by having them occur in the same decay volume simultaneously and almost collinearly. Unlike NA31 the $K_L^0$ beam was produced by a ten-times more energetic ($450~\mathrm{GeV}$) proton beam impinging on a beryllium target. The protons for the $K_S^0$ beam were produced by the free protons in the $K_L^0$ target that collide with a silicon monocrystal. Unlike NA31, a magnetic spectrometer was used for the measurement of charged decays with better accuracy. Although the generation of the decays $K_L^0 \to 2\pi$ required an intense initial beam of protons, a much less intense beam of protons was sufficient to create the $K_S^0$ beam when creating two-pion decays, which is approximately 5 orders of magnitude lower than the intensity of the beam for generating $K_L^0$s, which approximately corresponds to the square of the indirect violation parameter $|\eta^{+-}|$. 

The simultaneous generation of two kaon beams ensured their equal sensitivity to variations in the intensity of the primary proton beam within 10\%. Free protons (those not interacting with the target $K_L^0$) were
directed to a curved silicon crystal below the $K_L^0$ beam axis. This was a small fraction of protons ($\approx 10^{-5}$) due to its collision near the decay volume with the target $K_S^0$. The two-pion decays of this beam were decays due to the large difference in the lifetimes $K_L^0$ and $K_S^0$ and the low probability of such decays for $K_L^0$. The decay volume was $6~\mathrm{m}$ from the $K_S^0$ target. Here, the $K_S^0$ and $K_L^0$  beams exited from two apertures in the last collimator into the general decay region.  To identify decays in the beam $K_S^0$ a ``beam tagging station'' consisting of different scintillators was used. The triggers for recording decays $K^0\to 2\pi^0$ worked based on the analog summation of signals from the group of cells of a liquid krypton calorimeter.

The world average result is the following~\cite{Iconomidou-Fayard:2015lwg}:
\begin{equation}
    Re(\epsilon'/\epsilon) = (1.68\pm 0.20)\times 10^{-3}
     \>,
\nonumber
\end{equation}
\begin{equation}
    \frac{\Gamma (K^0\to \pi^+\pi^-) - \Gamma (\bar{K}^0\to \pi^+\pi^-)}
         {\Gamma (K^0\to \pi^+\pi^-) + \Gamma (\bar{K}^0\to \pi^+\pi^-)} = 2Re(\epsilon') = (5.3\pm 0.6)\times 10^{-6}
     \>,
\nonumber
\end{equation}
\begin{equation}
    \frac{\Gamma (K^0\to \pi^0\pi^0) - \Gamma (\bar{K}^0\to \pi^0\pi^0)}
         {\Gamma (K^0\to \pi^0\pi^0) + \Gamma (\bar{K}^0\to \pi^0\pi^0)} = -4Re(\epsilon') = (-10.6\pm 1.02)\times 10^{-6}
     \>.
\nonumber
\end{equation}
The recent data~\cite{ParticleDataGroup:2024pth} are given above in Section~\ref{Sec:CP-dir}.

\vspace{0.3cm}

Let us summarize the above. The NA48 experiment at CERN had made separate runs with long- and short-lived kaons. Both runs at CERN used simultaneous beams of long- and short-lived kaons and took data on charged- and neutral-pion production by both kaon beams at the same time. Installed on CERN’s highest intensity proton beamline, it uses a large and sophisticated detector. Some of the CERN protons go to make long-lived kaons, and the remaining particles are diffracted by a crystal and used to make a parallel beam of short-lived kaons. In this way, the two varieties of kaons are clearly differentiated, even though they eventually decay in the same way.
\begin{figure*}[htb!]
\centering
{
    \includegraphics[width=0.67\textwidth,keepaspectratio]{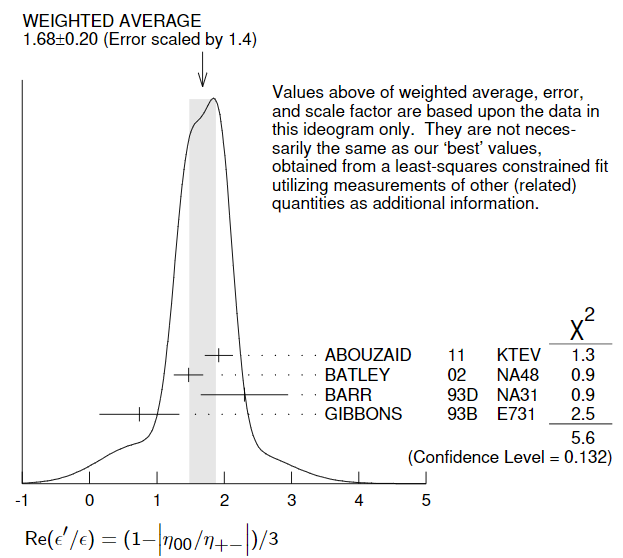}
}

\centerline{\parbox{0.86\textwidth}{
\caption[] {\protect\small
The comparison of the results of different experiments, taken from Ref.~\cite{ParticleDataGroup:2024pth}.
} 
\label{fig:fig5} } }
\end{figure*}

A detailed description of the selection of events in this and previous experiments is given in~\cite{Iconomidou-Fayard:2015lwg, Kekelidze:2007zz}. These reviews describe the beams, detectors, and analysis methods used in CERN experiments, which made significant contributions to their results and established new standards for precision measurements.

The final measurements of $\epsilon'/\epsilon$  made by the most precise experiments, KTEV, NA48, NA31, and E731, together with the average value are given in Fig.~\ref{fig:fig5}, taken from~\cite{ParticleDataGroup:2024pth}.  In contrast to the significant improvements of the NA48 experiment over the previous NA31 experiment at CERN, the KTeV experiment~\cite{KTeV:1999kad, KTeV:2002qqy} followed the approach of the previous E731 experiment, also at Fermilab. Both beams were formed by collision of $800~\mathrm{MeV}$  protons with a common production target; sweeping magnets removed the charged particles from the beams.  A common absorber increased the proportion of neutral kaons relative to neutrons in the beams. A plastic regenerator that promoted the formation of $K_S^0$ in one of the beams was moved to block the other neutral beam. The beams diverged from each other and were completely separated in the decay volume. The spectrometer, consisting of four drift chambers, including sensitive wires and a large dipole magnet, was filled with helium.  The electromagnetic calorimeter consisted of 3100 cells of cesium iodine (CsI) crystals. The trigger system included scintillation hodoscope signals for decays into charged pions and a summation of the energy release in the calorimeter during decays into neutral pions. The analysis of selected events included fitting the experimental data to curves obtained via Monte Carlo simulation of models for these processes. 

Using a subset of data collected in 1996--1997, the collaboration found~\cite{Hsiung:1999fk, Glazov:2000qv}: 
\begin{equation}
    Re(\epsilon'/\epsilon) = (2.80\pm 0.41)\times 10^{-3}
     \>.
\nonumber
\end{equation}
The final result of the KTeV experiment is it follows~\cite{KTeV:2010sng}:
\begin{equation}
    Re(\epsilon'/\epsilon) = (1.92\pm 0.21)\times 10^{-3}
     \>,
\nonumber
\end{equation}

A detailed description of the experiments NA31 and NA48 at CERN and KTeV at Fermilab is given in~\cite{Iconomidou-Fayard:2015lwg, Kekelidze:2007zz}. The data on $\epsilon'/\epsilon$  obtained in these experiments made it possible to unconditionally establish for the first time the presence of direct CP violation in the decays of neutral kaons. Given the most accurate result of the NA48 experiment (with the first indication of direct CP violation in experiment NA31), the contribution of other experiments is extremely important to a complete picture of the discovered phenomenon.  Detailed reviews are devoted to the violation of CP in neutral kaon decays~\cite{Popov:2021n, Iconomidou-Fayard:2015lwg, Kekelidze:2007zz, Lee:1965js, Okun:1982l, Bigi:2016i, Popov:2019n}.

In addition to obtaining new experimental results, theoretical research on the problem continued. The topics of Lorentz and CPT violation were discussed in Refs.~\cite{Colladay:1996iz, Kostelecky:2010ux, 
Cvetic:2000np}. They referenced the possibility of spontaneous CPT breaking that arises in string theory considering the SM extension (SME) taking into account this effect in the neutral meson system~\cite{Colladay:1996iz, Kostelecky:2010ux}. However, Lorenz symmetry breaking, although leading to the CPT violating terms, cannot lead to the CPT violation in experimentally interesting $K - \bar{K}$ and analogous systems~\cite{{Cvetic:2000np}}.  At the same time, any indication on CPT violation is absent up to now. 

\section{Alternative Research}
\label{Sec:AR}
\begin{figure*}[htb!]
\centering
{
    \includegraphics[width=0.7\textwidth,keepaspectratio]{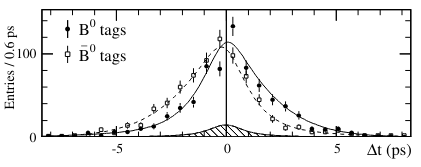}
}

\centerline{\parbox{0.86\textwidth}{
\caption[] {\protect\small
Measurement of $B^0\to J/\psi K^0$ decays from the BaBar Collaboration~\cite{BaBar:2002kla}. The word ``tags'' used here means that the initial-state tagging procedure was used to count the number of particles and antiparticles. The solid (dashed) curves represent the fit projection in $\Delta t$ for $B^0$ ($\bar{B}^0$) tags.
} 
\label{fig:fig6} } }
\end{figure*}

Beginning in 2001, a large number of CP violating processes in B-meson and D-meson decays have been discovered in B-factory experiments; these include the BaBar experiment at the Stanford Linear Accelerator Center (SLAC)~\cite{BaBar:2001ags}, the Belle Experiment at the High Energy Accelerator Research Organization (KEK)~\cite{Belle:2001qdd}, and the Large Hadron Collider Beauty Collaboration (LHCb) at CERN~\cite{LHCb:2013syl}. These experiments investigated the time-dependent CP asymmetry in the decays of $B^0\to J/\psi K^0$, where $J/\psi$ is a particle made of the charm and anti-charm quarks and was simultaneously discovered in 1974 by experimental groups at both SLAC and BNL confirming the existence of the charm quark $c$. Such final states provide a means to measure one of the angles ($\beta$) of the unitarity triangle (see above) with a negligibly small theoretical uncertainty. The difference between two distributions, \textit{i.e.}, $B^0$ and $\bar{B}^0$, in Figs.~\ref{fig:fig6} and 7
is a clearly visible indication that matter and antimatter are different.
\begin{figure*}[htb!]
\centering
{
    \includegraphics[width=0.7\textwidth,keepaspectratio]{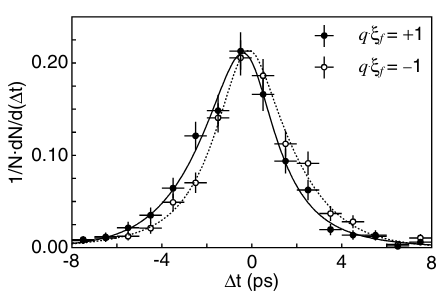}
}

\centerline{\parbox{0.86\textwidth}{
\caption[] {\protect\small
Measurement of $B^0\to J/\psi K^0$ decays from Belle Collaboration~\cite{Belle:2001zzw}. The parameter $q$ has discrete values $q = +1$ when the tag-side $B$-meson is more likely to be a $B^0$ and $q = -1$ when it is more likely to be a $\bar{B}^0$. So, they found 560 events with $q = +1$ flavor tags and 577 events with $q = -1$, observed $\Delta t$ distributions for $B^0$ (open circles) and $\bar{B}^0$ (filled circles).} 
}
\label{fig:fig7} } 
\end{figure*}

BaBar Collaboration obtained~\cite{BaBar:2001ags, BaBar:2001pki, BaBar:2002kla, BaBar:2004god}, respectively:
\begin{equation}
    \sin 2\beta = 0.34\pm 0.20(stat)\pm 0.05(syst)
     \>,
\nonumber
\end{equation}
\begin{equation}
    \sin 2\beta = 0.59\pm 0.14(stat)\pm 0.05(syst)
     \>,
\nonumber
\end{equation}
\begin{equation}
    \sin 2\beta = 0.741\pm 0.067(stat)\pm 0.034(syst)
     \>,
\nonumber
\end{equation}
\begin{equation}
    \sin 2\beta = 0.722\pm 0.040(stat)\pm 0.023(syst)
     \>.
\nonumber
\end{equation}

The study of decay $B^0\to J/\psi K$ is not the only possibility to obtain the angle $\beta$ with a small theoretical uncertainty. There are many $B$ meson decays with other final states that were also investigated by the Belle and BaBar Collaborations~\cite{Belle:2006dlp, BaBar:2021ich}. In the SM, other modes, such as $B^0\to J/\psi\phi$, measure the same angle, $\beta$, with comparable theoretical precision. The $\phi$ meson is a vector meson formed of a $s$- and $\bar{s}$-quarks. The existence of the $\phi$ meson was first proposed by Sakurai in 1962 as a resonance state between the $K^0$ and $\bar{K}^0$~\cite{Sakurai:1962zz}. It was discovered later by Connolly \textit{et al.} at the Alternating Gradient Synchrotron (AGS) at BNL~\cite{Connolly:1963pb}. 

The consideration via a loop transition, \textit{i.e.}, penguin diagrams, is very sensitive to possible contributions from unknown particles, provided there are probes for new sources of CP violation~\cite{Barel:2022wfr}. The measurement of CP violation in pinging modes is experimentally very desirable, due to the extremely low rates of the final states of interest and the difficulty in separating these from various sources of background.  The favorable conditions of the $B$ factories (\textit{e.g.} high luminosity \textit{etc.}) make them ideal laboratories in which to perform these measurements. The data obtained show that the angle $\beta$ in the penguin diagram differs by $2.7\sigma$ compared to the measurement of the same angle in the SM. However, testing if the modern data agree with the SM prediction, the research for penguin-dominated modes continues. The $b\to s\bar{q}q$  (where $q = b$, $s$, $d$) penguin decays have the same CKM phase as the decays at tree level $b\to s\bar{c}c$ suppressed by $\lambda^2$. Therefore, decays such as $B^0\to \phi K^0$ provide $\sin 2\beta$ measurements in the SM. The main interest in these modes is not simply to measure $\sin 2\beta$, but to search for new physics beyond SM (BSM). The mass of these particles that appear virtually is expected to exceed hundreds of GeV, and either their decay or creation by the method of lost energy is sought. 

The registration of CP violation in charm decays was also performed by the LHCb Collaboration. There are many other experiments sensitive to the BSM supposition, including a $B_s$ system with a small angle.
\begin{equation}
    \beta_s = arg\Bigl(-\frac{V_{ts}V_{tb}^\ast}{V_{cs}V_{cb}^\ast}\Bigl)
     \>.
\nonumber
\end{equation}
This angle can be  measured via time-dependent CP violation in $B^0_s\to J/\psi \phi$ analogically to $\beta$ in $B^0\to J/\psi K^0$.  Since the $J/\psi \phi$ final state is not a CP eigenstate, an angular analysis of the decay products was performed to separate the CP-even and CP-odd components, which gave opposite asymmetries. The comparison of the LHCb results, including $B_s\to J/\psi K^+K^-$ and $B_s\to J/\psi \pi^+\pi^-$, for
$$2\beta_s = 0.039\pm 0.022(stat) \pm 0.006(syst)$$
and the current world average,   
$$2\beta_s = 0.040\pm 0.016$$
with the SM prediction   
$$2\beta_s = 0.03726^{+0.00078}_{-0.00077}$$
indicates an excellent agreement. LHCb also measured~\cite{LHCb:2023zcp}:
$$\sin2\beta = 0.717\pm 0.013(stat)\pm 0.008(syst) \>.$$

References~\cite{LHCb:2019jta, LHCb:2019sus} indicate that CP parity violation is evident in the ratio of measured yields for the decays of the $B^+$ and $B^-$ mesons, which are a particle and antiparticle pair. The decay of the $B^-$ meson through the $3\pi$ channel is more intense than that of the $B^+$ meson. During these decays, large amounts of energy are released owing to the heavy mass of the B mesons in comparison to the relatively light pions; the decay energy is carried away predominantly by the kinetic energy of the pions. These decays play an important role in the production of intermediate resonances that decay into pion pairs with a lifetime of the order of $10^{-24}~\mathrm{sec}$.  Concurrently, the $V_{ub}$ element of the CKM matrix responsible for direct $b\to u$ transitions is very small, so the contribution from the corresponding tree amplitudes is comparable to the contribution of loop diagrams (such as the penguin diagram). Tree and loop diagrams interfere (amplifying or compensating each other), depending on the imaginary part of the CKM matrix, which has a different sign for particles and antiparticles.

The effect of direct CP violation can manifest itself more strongly if one observes not just the integral number of decays but the difference at a number of discreet points of the phase volume. Dalitz diagrams have been constructed for the two-dimensional distribution of $m^2(\pi^+\pi^-)$  pairs of daughter particles, on which resonance bands are visible. It can be seen that the nature of the intermediate states is different for the $B^+$ and $B^-$ mesons and in some areas of these diagrams the CP asymmetry, that is, the ratio of the difference in the yields $B^+$ and $B^-$ to their sum, reaches 80\% with the prevalence of the yield of $B^-$ mesons.

The issue of counting the number of particles and antiparticles is solved by the initial state tagging procedure.  To register CP violation when mixing neutral mesons, it is necessary to establish the flavor of the particle that was born initially; \textit{i.e.} was it a particle or an antiparticle?  The period of quark-antiquark oscillations is small ($\approx \mathrm{nsec}$) compared to the lifetime of a meson.

CP violation was also predicted for $D$ mesons, which includes a representative of the second type of quark with a charge of $+2/3$, \textit{i.e.}, a charmed $c$-quark~\cite{Feldmann:2012js}. The latest publication of the LHCb Collaboration reports that it was possible for the first time to establish accurately the CP violation for $D$ mesons. The asymmetry $A$ of the decay of the $D$ meson in relation to $\bar{D}$ was studied in two different channels, in $\pi^+\pi^-$ and $K^+K^-$. The difference between these depends on the unequal determination of the decay products of the two channels; the difference in asymmetries in these decay channels reduces the detection effect, leaving only CP violation.  The result was symmetry violation~\cite{LHCb:2019hro}:
\begin{equation}
    \Delta A_{CP} = A_{CP}(D^0\to K^-K^+) - A_{CP}(D^0\to \pi^-\pi^+)
     \>
\nonumber
\end{equation}
is
\begin{equation}
    \Delta A_{CP} = (-18.2\pm 3.2(stat)\pm 0.9(syst))\times 10^{-4}
     \>
\nonumber
\end{equation}
for $\pi$ tagged $D^0$ meson in the reaction $D^\ast(2010)^+\to D^0 \pi^+$.
$$\Delta A_{CP} = (-9\pm 8(stat)\pm 5(syst))\times 10^{-4},$$
for $\mu$ tagged $D^0$ meson in the reaction $\bar{B}\to D^0 \mu^- \bar{\nu_\mu}X$.

Combining these with previous LHCb results leaded to
\begin{equation}
    \Delta A_{CP} = (-15.4\pm 2.9)\times 10^{-4}
     \>.
\nonumber
\end{equation}
Where the uncertainty includes both statistical and systematic contributions. It differs from zero by more than $5\sigma$.

A long-term study of the rare decay of the $B$ meson into a lepton pair has established a violation of lepton invariance favoring its decay into an electron-positron pair rather than decay into a $\mu^+\mu^-$  pair. This is not consistent with universality within the SM framework; the experimental ratio of the decay probabilities is $R_k = (B^0\to \mu^+\mu^-)/(B^0\to e^+e^-) = 0.846\pm 0.044$ (LHCb) compared to the expectation of unity. At the same time, the values of the decay rate of $B^0$ in a pair of muons $B_s\to \mu^+\mu^-$ were calculated within the SM framework:
\begin{equation}
    B_{th}(B_s) = (3.56\pm 0.30)\times 10^{-9}
     \>,
\nonumber
\end{equation}
\begin{equation}
    B_{th}(B) = (0.10\pm 0.01)\times 10^{-9}
     \>.
\nonumber
\end{equation}
The experimental value obtained by the LHCb collaboration from Ref.~\cite{LHCb:2021vsc} is equal to 
\begin{equation}
    B(B^0_s) = (3.09^{+0.46+0.15}_{-0.13-0.11})\times 10^{-9}
     \>.
\nonumber
\end{equation}
No significant signal for $B^0\to \mu^+\mu^-$ decay was found with an upper limit of $B(B^0) < 0.26\times 10^{-9}$.
The world average~\cite{ParticleDataGroup:2024pth}:
\begin{equation}
    B_{exp}(B_s) \to (3.01\pm 0.35)\times 10^{-9} 
     \>
\nonumber
\end{equation}
is consistent with the SM, with the current uncertainties.

The rare  decay ($K_S\to \mu^+\mu^-$)  was investigated in Ref.~\cite{Brod:2022khx} as a sensitive probe of the short-distance region.  SM predictions for the short-distance contribution improved by a few percent. As the CP violating effect is small, $\approx 10^{-3}$ any other corrections should be taken into account to assure accurate extraction $\epsilon$ and $\epsilon'$ from the experimental data.

Reference~\cite{Buccella:1994pj} considered the CP violating asymmetry in the energy of the charged pions in the decay $K_L^0\to \pi^+\pi^-\pi^0$ in terms of two parameters that were determined by studying the corresponding time-dependent asymmetry in the decay of the neutral kaons with fixed initial strangeness. The measure of the conserving CP asymmetry provides a test of Chiral Perturbation Theory ($\chi$PT) and gives information on the final-state interaction of the three pions.

Fleischer provided a review~\cite{Fleischer:2024uru} on CP violation in B decays with respect to recent development and future perspectives. He indicated that these processes provide a powerful tool to probe new physics beyond the Standard Model (BSM). The review of recent data on the decays of $B$ is also given by Malami in~\cite{Malami:2024ptc}.

Higher-order electroweak contributions to indirect CP violation in neutral kaons were calculated in~\cite{Brod:2024sgj}. The authors wrote that the theoretical calculation of $\epsilon_K$ had historically large perturbative errors arising from charm-quark corrections. These errors were larger than expected higher-order electroweak corrections. Recently, a simple reparameterization of the effective Hamiltonian reduced perturbation errors, making higher-order electroweak calculations justified. They presented the leading logarithmic electroweak contribution of O(1\%) to $\epsilon_K$.

Starting in 2012 the KOTO Collaboration using the J-PARC $30~\mathrm{GeV}$ Main Ring at KEK, investigated rare decay $K_L\to \pi^0\nu\bar{\nu}$ due to direct CP violation~\cite{KOTO:2020prk}. They obtained the upper limit of 
\begin{equation}
    BR(K_L\to \pi^0\nu\bar{\nu}) < 5.1\times 10^{-8}~(90\%~\mathrm{C.L.}) 
     \>.
\nonumber
\end{equation}
This mode directly breaks CP symmetry and is highly suppressed in the SM. Moreover, the theoretical uncertainty of this decay is only a few percent; these features make this decay one of the best probes to search for BSM physics. They concluded that the number of observed events is statistically consistent with the background expectation. The agreement of theory in the SM framework with this experimental result could be an indication that BSM physics is unnecessary here.

An analogous search was undertaken by the NA62 Collaboration at CERN to perform the first search for the rare $K^+\to \pi^+ \nu \bar{\nu}$ decay channel~\cite{NA62:2018ctf}.  They obtained an upper limit of
\begin{equation}
    BR_{exp}(K^+\to \pi^+\nu\bar{\nu}) < 14\times 10^{-10}~(95\%~\mathrm{C.L.}) 
     \>.
\nonumber
\end{equation}
This value agrees, within uncertainties, with the prediction of the SM:
\begin{equation}
    BR_{theor}(K^+\to \pi^+\nu\bar{\nu}) < 10\times 10^{-10}~(95\%~\mathrm{C.L.}) 
     \>,
\nonumber
\end{equation}
ruling out any need to resort to BSM theories. The kaons were created by a $450~\mathrm{GeV}$ proton beam from the CERN SPS that impinged on a beryllium production target.  A 270-meter-long detector, using a combination of triggers, made it possible to select kaon decays needed to calculate the background and search for the desired decay channel. The current theoretical and experimental conditions allowed for the first detection of the “golden modes" of this ultra-rare K decay channel with a statistical uncertainty of $3.5\sigma$.  The current theoretical and experimental conditions concerning the 'golden modes' of $K\to \pi\nu\bar{\nu}$, are taken from~\cite{ParticleDataGroup:2024pth} and are summarized in Fig.~8.

\begin{figure*}[htb!]
\centering
{
    \includegraphics[width=0.7\textwidth,keepaspectratio]{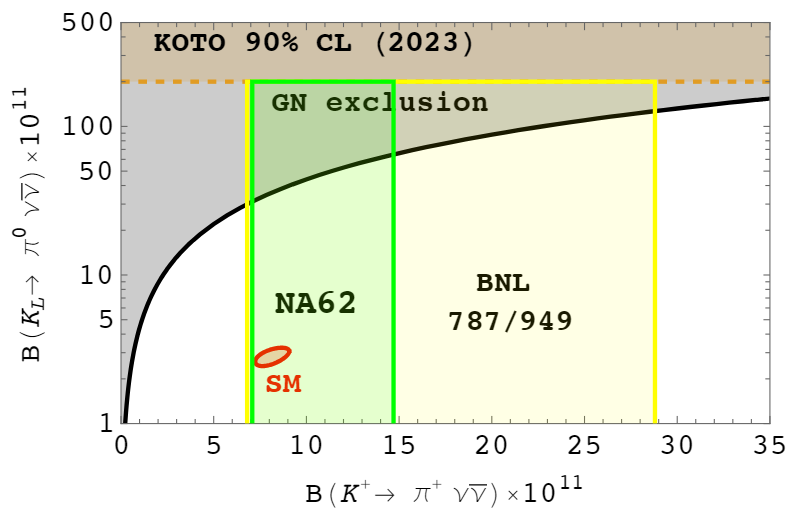}
}

\centerline{\parbox{0.86\textwidth}{
\caption[] {\protect\small
Summary of current situation for the “golden modes" $K\to \pi \nu \bar{\nu}$, taken from Ref.~\cite{ParticleDataGroup:2024pth}.  The red ellipse shows the prediction $1\sigma$ of SM with input from the CKM fitter~\cite{Charles:2011va}; the green (yellow) region corresponds to the NA62 (BNL787/949)\ 1$\sigma$ measurement; and the dashed orange line marks the upper bound 90\%~C.L. KOTO~\cite{KOTO:2018dsc}. The black shaded region shows the Grossman and Nir~\cite{Grossman:1997sk} exclusion.}
} 
\label{fig:fig8} } 
\end{figure*}

The world’s leading program for the decay of K mesons is HIKE~\cite{Anzivino:2023bhp}.  The program investigates several ultra-rare ``golden'' kaon decays, which are very clean from the theory of SM. The measurement of $BR(K^+\to \pi^+ \nu\bar{\nu})$ with 5\% precision (corresponding to the accuracy of the SM calculations) refers to the phase $1$ of the program, while the phase $2$ corresponds to the first observation of decays of $K^+\to \pi^0 l^+l^-$ with a precision exceeding 5\%.  Being essentially free of hadronic uncertainties, they allow for precise tests of the SM. Their suppression in the SM leads to sensitivity to new BSM physics.   HIKE will use high-intensity $K^+$ and $K_L$ beams and collect the world's largest decay samples $K^+$ and $K_L$, using triggers and detectors that are better than NA62 and other previous experiments. As a result, HIKE will significantly improve the previous of measurements. In addition to the rare gold modes, HIKE and KOTO-II will measure many other $K^+$ and $K_L$ decays, including radiative ones.with high precision.

The CP violating processes observed so far fit well into the framework of the SM.  However, surprises cannot be ruled out when studying such processes with the need to search for alternative sources of CP violation with access to new physics. This problem is discussed in~\cite{Grossman:1997sk}. A measurement of time-dependent CP violation in the decays of the $B^0$ and $\bar{B}^0$ mesons to the final states $J/\psi (\to \mu^+ \mu^-)K_S^0$, $\psi(2S)(\to \mu^+ \mu^-)K_S^0$, and $J/\psi (\to e^+e^-)K_S^0$ with $K_S^0\to \pi^+\pi^-$ (the $J/\psi$ meson was discovered in 1974 by two groups of experimentalists at SLAC and BNL, confirming the existence of charm quark $c$) was considered in Ref.~\cite{LHCb:2023zcp}. In agreement with their statement, the CP violation parameters obtained represent the most prestigious single measurement of the CKM angle $\beta$ to date and were more precise than the current global average.

The existing experiments indicate that the CP violation is too weak compared to what is required to explain the matter-antimatter asymmetry. The magnitude of the CP violation effect will be proportional to the function of all three mixing angles and the $\delta$ known as the Jarlskog invariant~\cite{ParticleDataGroup:2024pth}: 
\begin{equation}
    J_{CP}^{CKM} = c_{12}~c_{13}^2~c_{23}~s_{12}~s_{13}~s_{23}~\sin\delta = (3.12^{+0.13}_{-0.12})\times 10^{-5}
     \>.
\nonumber
\end{equation}
The quark sector cannot explain the baryon asymmetry of the Universe while the mixing angles are small. An analysis of this asymmetry, $\Delta A_{CP}$, taking into account the direct CP violations discovered in $D^0\to K^-K^+$ and $D^0\to \pi^-\pi^+$ decays~\cite{LHCb:2022lry} and searching for CP violation in the radiative charm decays of the $D^0$ meson,
is given in~\cite{Chernov:2024nrx}. The recent results are reported at ``the 2024 International Workshop on Future Tau Charm Facilities,'' Shanghai, China, 14-18 January 2024,
https://indico.pnp.ustc.edu.cn/event/91/ and at ``22nd Conference on Flavor Physics and CP Violation,'' Bangkok, Thailand, 27-31 May 2024, https://indico.cern.ch/event/1291023/.

An open question remains to understand the participation of heavy quarks during CP violation in the weak interaction, whereas spatial parity violation ($P$-violation) is characteristic of all weak interaction processes. In contrast to $P$ parity violation in weak interaction, CP parity violation does not exceed a fraction of a percent. Finally, further explaining the CP violation BSM, that is, within the framework of new physics, is an important task in studies of this problem~\cite{Nir:1995nw, Gershon:2022, Martinelli:2024esg}. One of the fundamental problems of modern physics is the explanation of the source of CP violation in the weak interaction. The first evidence for direct CP violation in beauty to charmonium decays was given in Ref.~\cite{LHCb:2024exp}.

\section*{Acknowledgments}
One of us, N.P., thanks Dr.~Wilhelm~Czaplinski for collaboration. 
I.S. thanks Dr.~Valery Kubarovsky for useful comments.
This work was supported in part by the U.~S.~Department of Energy, Office of Science, Office of Nuclear Physics, under Award No.~DE--SC0016583.


\end{document}